\documentclass{Interspeech2024}

\usepackage{graphicx}
\usepackage{caption}
\usepackage{subcaption}

\interspeechcameraready

\title{A Toolkit for Joint Speaker Diarization and Identification with Application to Speaker-Attributed ASR}

\name[affiliation={1}]{Giovanni}{Morrone}
\name[affiliation={1}]{Enrico}{Zovato}
\name[affiliation={1}]{Fabio}{Brugnara}
\name[affiliation={1}]{Enrico}{Sartori}
\name[affiliation={1}]{Leonardo}{Badino}

\address{
  $^1$Almawave S.p.A., Voice Engineering Lab, Italy}
\email{\{g.morrone,e.zovato,f.brugnara,e.sartori,l.badino\}@almawave.it}


\begin{document}

\maketitle

\begin{abstract}
    We present a modular toolkit to perform joint speaker diarization and speaker identification. The toolkit can leverage on multiple models and algorithms which are defined in a configuration file. Such flexibility allows our system to work properly in various conditions (e.g., multiple registered speakers' sets, acoustic conditions and languages) and across application domains (e.g. media monitoring, institutional, speech analytics). In this demonstration we show a practical use-case in which speaker-related information is used jointly with automatic speech recognition engines to generate speaker-attributed transcriptions. To achieve that, we employ a user-friendly web-based interface to process audio and video inputs with the chosen configuration. 
\end{abstract}

\section{Introduction}
\label{sec:intro}

    
Speaker identification consists in assigning real person identities to input audio segments. Speaker identity is a valuable information which is useful for many applications, such as speaker-attributed automatic speech recognition (ASR), speaker-based indexing, voice biometrics and more.

    
In this demonstration we address the problem of joint speaker diarization (SD) and speaker identification (SI). Contrary to speaker verification, there is no a priori identity claim, and the system determines who the person is among a fixed set of registered speakers. Additionally, in the open-set case the system can decide that the person is unknown. We show a practical use-case in which both SD and SI information are exploited to generate speaker-attributed transcriptions.

    
    
Although several open-source and commercial products are available, they are intended for people with programming skills. Typically, only general purpose models and systems are available for end users. Such solutions sometimes fail for specific application domains, making them of limited use in real use-cases.
We aim at making these technologies available also for users without expertise in the field. We employ a user-friendly web-based application which allows users to choose among different setups that reflect various application domains (e.g. media monitoring, speech analytics, institutional). A setup defines SD and SI models, alongside with the ASR engine used for generating transcriptions. 
    
    
Such flexibility increases overall system robustness across different conditions (e.g. speech styles, languages, signal quality, speakers' identities and characteristics), while maintaining an easy-to-use interface for the end user. These features are very relevant for the Interspeech 2024 theme.

\begin{figure}[t]
    \centering
    \hspace*{-5pt}
    \includegraphics[width=0.42\textwidth]{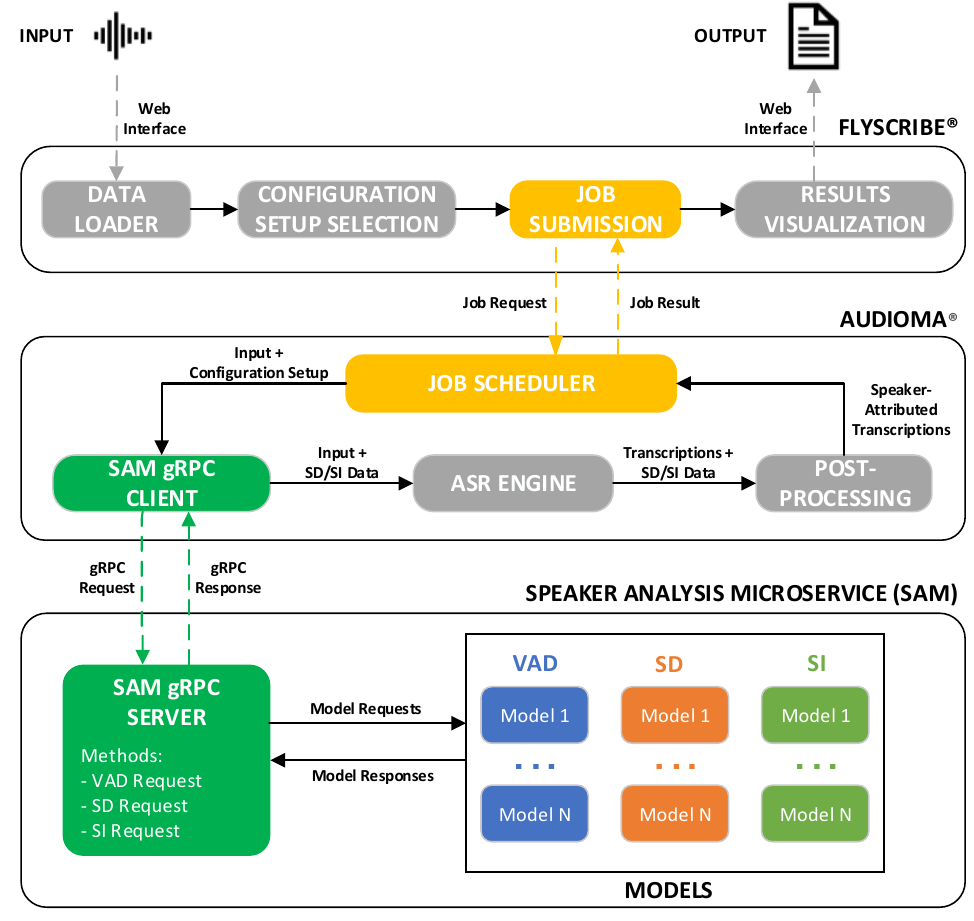}
    \vspace{-5pt}
    \caption{Block diagram of the overall system architecture.}
    \vspace{-15pt}
    \label{fig:diagram}
\end{figure}

\section{System Architecture}
\label{sec:system}

Figure \ref{fig:diagram} shows a diagram of the proposed system architecture. It consists of three building blocks:

\begin{enumerate}
   \item Speaker Analysis Microservice (SAM): it provides the speaker-related functions (e.g., SD and SI). 
   \item Audioma\textsuperscript{\textregistered}: it controls job scheduling and orchestration. In particular it manages requests to the SAM and controls processing of ASR engines.
   \item FlyScribe\textsuperscript{\textregistered}: graphical user interface for Audioma\textsuperscript{\textregistered}.
\end{enumerate}

\vspace{-5pt}
\subsection{Speaker Analysis Microservice (SAM)}
\label{ssec:microservice}

The SAM is based on the gRPC framework\footnote{\url{https://grpc.io/}} to efficiently provide speaker-related services to external servers and programs. It supports three functions: SD, voice activity detection (VAD) and SI.

Our SD models follow the end-to-end neural diarization (EEND)-vector clustering framework \cite{kinoshita21_interspeech}. It is a hybrid diarization approach which combines neural and clustering-based diarization. In addition to the original implementation, we have designed an improved version which reaches better results for long input signals with many speakers (e.g., $>5$). In particular, it replaces the speaker embeddings generated by neural diarization with those obtained from an external pre-trained embedding extractor. We use speaker embedding extractors available in the Wespeaker toolkit\footnote{\url{https://github.com/wenet-e2e/wespeaker/tree/master/examples/voxceleb/v2}} \cite{wang2023wespeaker} and trained on VoxCeleb2 \cite{chung2018voxceleb2}.

The SAM also allows to only use VAD derived from SD models. In that case, speaker-related information is discarded and only speech/non-speech segmentation is retained. Indeed, if SD is not needed then speaker embeddings extraction and clustering can be avoided to save computations. 

Our SI system is based on speaker embedding extractors. As for SD, we employ the Wespeaker toolkit. Given an input segment, it determines whether it is uttered by a speaker belonging to a fixed set of registered speakers or by an unknown speaker. In a preliminary enrollment stage, a reference speaker embedding is computed for each registered speaker from the fixed set. During inference, a speaker embedding is extracted from the input segment and a similarity score is computed for each of the reference speaker embeddings of the fixed set. In the close-set case, the system returns the person's identity with the highest score. In the open-set case, the system can mark the input as unknown if all scores are below a similarity threshold. In this case, numeric labels are used to distinguish between different unknown speakers (e.g., Speaker1, \ldots, SpeakerN). SI inference is performed for each speech segment detected by SD or VAD. Additionally, SI can leverage SD information about speaker identities and speech overlap to derive better speaker embeddings. 

\subsection{Audioma\textsuperscript{\textregistered}}
\label{ssec:audioma}

Audioma\textsuperscript{\textregistered} is an in-house product which provides several speech-related functions and features, including utilities for sending requests to other services (e.g., SAM), execution of ASR engines and job orchestration (i.e., submission, queuing, parallelization, processing). The whole process can be controlled by setting one or more options in specific configuration files. In particular, the most relevant options for this demonstration are the following:

\begin{itemize}
    \item Preprocessing step: SD or VAD model. 
    \item SI model and registered speakers' set selection.
    \item ASR engine: multiple languages and domains are available.
\end{itemize}

Both general-purpose and domain-specific (e.g., media monitoring, speech analytics) modules are available. The best configuration often depends on the final application.


\subsection{FlyScribe\textsuperscript{\textregistered}}
\label{ssec:flyscribe}

FlyScribe\textsuperscript{\textregistered} is a web application that provides a graphical user interface for Audioma\textsuperscript{\textregistered} functions. Among others, it includes transcription, automatic subtitling and machine translation services.

In particular, in this demonstration we use the transcription service. Three main steps are involved: configuration setup selection, input submission, results visualization. Two screen captures of the user interface are shown in Fig. \ref{fig:screenshot}.

\section{Demo Application}
\label{sec:demo}

In this section we briefly describe what participants are expected to see during the demonstration.

\begin{figure}[t!]
\centering
    \begin{subfigure}[b]{0.27\textwidth}
        \includegraphics[width=1.0\linewidth]{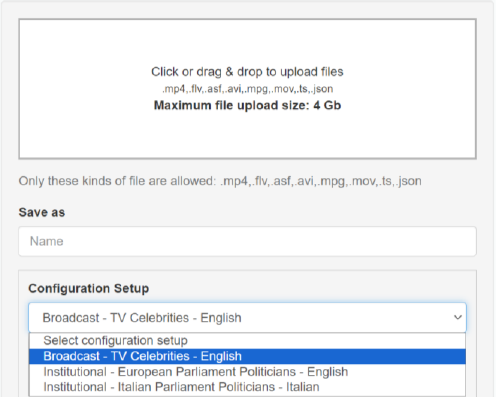}
        \vspace{-18pt}
        \caption{Submission}
        \label{fig:screen_submission}
        \vspace{5pt}
    \end{subfigure}
    \begin{subfigure}[b]{0.42\textwidth}
        \includegraphics[width=1.0\linewidth]
        {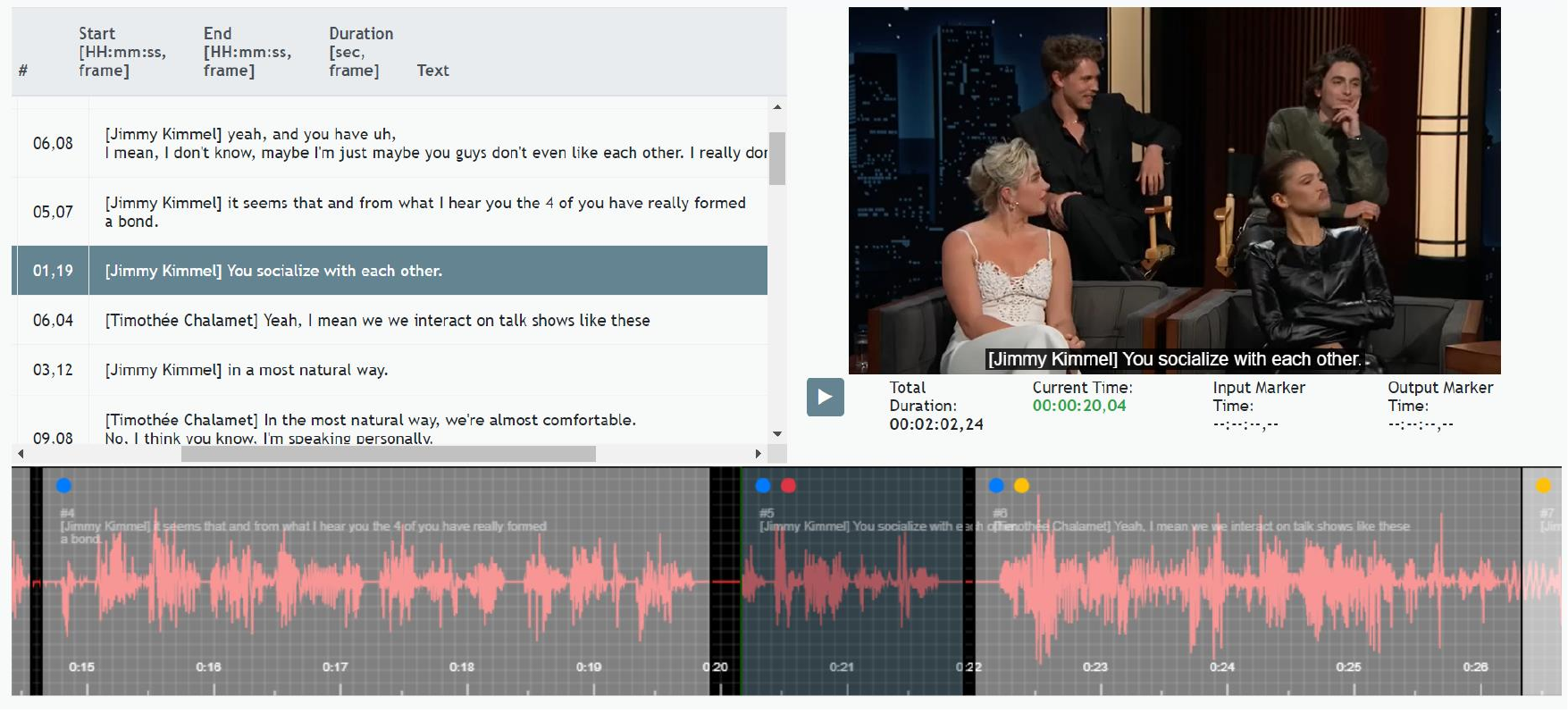}
        \vspace{-15pt}
        \caption{Results visualization}
        \label{fig:screen_results}
        \vspace{-5pt}
    \end{subfigure}
    \caption{Screen captures of the demo graphical user interface.}
    \label{fig:screenshot}
    \vspace{-15pt}
\end{figure}

Firstly, the user can choose between different pre-compiled configuration setups (see Figure \ref{fig:screen_submission}). We provide both general and domain-specific setups. For each setup a detailed description is provided to help users select the most suitable configuration for their use-cases. Then, the user can upload the audio or video and submit the request to the system.


When the submitted job is completed the user can visualize the speaker-attributed transcriptions results in a dedicated web interface (see Figure \ref{fig:screen_results}). Additionally, an audio/video player with synchronized annotations is available. This feature is useful to allow users to navigate long recordings as users may only be interested in specific sections. The annotations can be manually edited, if needed, to guarantee human-level accuracy and readability. Finally, the results can be exported in standard formats (e.g., SRT for subtitle files).

We employ a Linux-based server with an Intel\textsuperscript{\textregistered} Core\textsuperscript{\texttrademark} i7-9800X CPU @ 3.80GHz using one CPU thread without any GPU. We achieve a real-time factor of 0.18 to process a transcription job. That is acceptable for an interactive demonstration.

\vspace{-8pt}
\section{Conclusions}
\label{sec:conclusions}

Speaker diarization and identification technologies have reached outstanding performance in the last few years. However, the joint SD and SI task has not gained much attention. Moreover, such technologies are not easily available for end users with no programming skills. To bridge this gap, we develop a toolkit for joint SD and SI which efficiently combines the latest speaker-related technologies to guarantee both high accuracy and inference speed in different application domains. We propose an interactive demonstration in which the proposed system is used along with ASR engines to generate speaker-attributed transcriptions. Through an user-friendly interface users can choose the configuration setup most suitable to their needs, visualize results and export them in standard human-readable formats.

\vspace{-4pt}
\bibliographystyle{IEEEtran}
\bibliography{mybib}

\begin{thebibliography}{1}
\providecommand{\url}[1]{#1}
\csname url@samestyle\endcsname
\providecommand{\newblock}{\relax}
\providecommand{\bibinfo}[2]{#2}
\providecommand{\BIBentrySTDinterwordspacing}{\spaceskip=0pt\relax}
\providecommand{\BIBentryALTinterwordstretchfactor}{4}
\providecommand{\BIBentryALTinterwordspacing}{\spaceskip=\fontdimen2\font plus
\BIBentryALTinterwordstretchfactor\fontdimen3\font minus \fontdimen4\font\relax}
\providecommand{\BIBforeignlanguage}[2]{{%
\expandafter\ifx\csname l@#1\endcsname\relax
\typeout{** WARNING: IEEEtran.bst: No hyphenation pattern has been}%
\typeout{** loaded for the language `#1'. Using the pattern for}%
\typeout{** the default language instead.}%
\else
\language=\csname l@#1\endcsname
\fi
#2}}
\providecommand{\BIBdecl}{\relax}
\BIBdecl

\bibitem{kinoshita21_interspeech}
K.~Kinoshita, M.~Delcroix, and N.~Tawara, ``{Advances in Integration of End-to-End Neural and Clustering-Based Diarization for Real Conversational Speech},'' in \emph{Proc. of Interspeech}.\hskip 1em plus 0.5em minus 0.4em\relax ISCA, 2021, pp. 3565--3569.

\bibitem{wang2023wespeaker}
H.~Wang, C.~Liang, S.~Wang, Z.~Chen, B.~Zhang, X.~Xiang, Y.~Deng, and Y.~Qian, ``Wespeaker: A research and production oriented speaker embedding learning toolkit,'' in \emph{Proc. of International Conference on Acoustics, Speech and Signal Processing}.\hskip 1em plus 0.5em minus 0.4em\relax IEEE, 2023, pp. 1--5.

\bibitem{chung2018voxceleb2}
J.~S. Chung, A.~Nagrani, and A.~Zisserman, ``{VoxCeleb2: Deep speaker recognition},'' in \emph{Proc. of Interspeech}.\hskip 1em plus 0.5em minus 0.4em\relax ISCA, 2018.

\end{thebibliography}

\end{document}